 \documentclass[12pt,preprint]{aastex}

\def\gsim{\;\lower4pt\hbox{${\buildrel\displaystyle >\over\sim}$}\;}
\def\lsim{\;\lower4pt\hbox{${\buildrel\displaystyle <\over\sim}$}\;}

\shorttitle{Relativistic corrections on SZE}
\shortauthors{Fan \& Wu}

\begin{document}


\title{The Effects of Relativistic Corrections on Cosmological Parameter Estimations
from SZE Cluster Surveys}


\author{Zuhui Fan\altaffilmark{1,2} and Yanling Wu\altaffilmark{1}}



\altaffiltext{1}{Department of Astronomy, Peking University,
Beijing 100871, China}
\altaffiltext{2}{Beijing Astrophysics Center, Chinese Academy of Science and
Peking University, Beijing 100871, China }


\begin{abstract}
Sunyaev-Zel'dovich Effect (SZE) cluster surveys are anticipated to yield
tight constraints on cosmological parameters such as the equation of
state of dark energy. In this paper, we study the impact of relativistic
corrections of the thermal SZE on the cluster number counts expected from
a cosmological model and thus, assuming that other cosmological parameters are known
to high accuracies, on the determination of the $w$ parameter 
and $\sigma_8$ from a SZE cluster survey, where $w=p/\rho$ with
$p$ the pressure and $\rho$ the density of dark energy, and $\sigma_8$
is the rms of the extrapolated linear density fluctuation 
smoothed over $8\hbox{ Mpc}h^{-1}$. 
For the purpose of illustrating the effects of relativistic corrections,
our analyses mainly focus on $\nu=353 \hbox{ GHz}$ and
$S_{lim}=30\hbox{ mJy}$, where $\nu$ and $S_{lim}$ are the observing frequency 
and the flux limit of a survey, respectively. These observing parameters are relevant to
the {\it Planck } survey.
It is found that from two measurable quantities, the total number of
SZE clusters and the number of clusters with redshift $z\ge 0.5$,
$\sigma_8$ and $w$ can be determined to a level of $\pm 1\%$ and
$\pm 8\%$, respectively, with $1\sigma$ uncertainties 
from a survey of $10000\hbox{ deg}^2$. Relativistic
effects are important in determining the central values of 
$\sigma_8$ and $w$. If we choose the two quantities calculated 
relativistically from the flat cosmological model with $\sigma_8=0.8284$ and $w=-0.75$ 
as input, the derived $\sigma_8$ and $w$ would be $0.819$ and $-0.81$, 
respectively, if relativistic effects are wrongly neglected. 
The location of the resulting $\sigma_8$ and $w$ in the $\sigma_8-w$ plane is outside 
the $3\sigma$ region around the real central $\sigma_8$ and $w$. 

\end{abstract}




\section{Introduction}

With fast advances in astronomical observations, the determination
of cosmological parameters has become one of the most important studies
in cosmology (e.g., Eisenstein, Hu \& Tegmark 1999; Spergel et al. 2003). 
Different kinds of observables are sensitive to 
different underlying physical processes, and thus can be used to probe different cosmological 
parameters. The local abundance and the evolution of clusters of
galaxies have been used extensively in the $\Omega_m$ and $\sigma_8$
determination, where $\Omega_m$ is the present matter density parameter of
the universe and $\sigma_8$ is the rms of the extrapolated linear density fluctuation
smoothed over $8\hbox { Mpc}h^{-1}$ (e.g., Bahcall \& Fan 1998; Pen 1998; 
Haiman, Mohr \& Holder 2001; Holder, Haiman \& Mohr 2001; Fan \& Chiueh 2001; 
Rosati, Borgani \& Norman 2002). 
Supernova type Ia (SNeIa) (Schmidt  et al. 1998; Perlmutter et al. 1999) 
and Cosmic Microwave Background (CMB) radiation observations (e.g., Spergel et al. 2003)
provide convincing evidences that we are living in a flat universe with 
about $70\%$ of the matter being in the form of dark energy. 
As the properties of dark energy affect the global geometry
of the universe and the growth of matter density fluctuations,
the abundance and the evolution of clusters are dependent of 
the $w$ parameter of the equation of state of dark energy, where
$w=p/\rho$ with $p$ the pressure and $\rho$ the density of dark energy
(e.g., Caldwell, Dave, \& Steinhardt 1998).
Studies show that deep and large cluster surveys could constrain 
$w$ to about $5\%$ level (e.g., Mohr et al. 2002). 
In order to reach such a precision, however,
we have to understand different types of systematic effects.
Investigations indicate that the incompleteness of the knowledge about the intracluster 
gas (ICM) affects the cosmological parameter determination from X-ray cluster
surveys, but influences less on the results from Sunyaev-Zel'dovich Effect (SZE) 
cluster surveys (Majumdar \& Mohr 2003). In this paper, we study the impact of
relativistic corrections of thermal SZE on the $w$ and $\sigma_8$ estimation from 
SZE cluster surveys. 

The typical gas temperature in clusters of galaxies is about
a few kev, but clusters with temperatures as high as 
about $17\hbox{ kev}$ have been observed (e.g., Pointecouteau et al. 1999; 2001).
Thus relativistic corrections on SZE can be significant (e.g., Birkinshaw 1999; 
Carlstrom, Holder \& Reese 2002; Rephaeli 2002). 
For a SZE cluster survey,
the selection function is mostly related to a mass threshold
that depends on the sensitivity of the instrument and is also
a function of redshift. For a survey with a relatively high
flux limit, only clusters with large masses, and therefore high
temperatures, can be detected. In this case, we expect that 
relativistic corrections will impact the theoretical
estimations of the survey results considerably. The amplitude of the change of SZE due to
relativistic effects also depends
on observing frequencies, with a relatively large correction 
around $400\hbox{ GHz}$. For {\it Planck},
one of the best frequencies for SZE observation is $353\hbox{ GHz}$ and the 
flux limit is about $30\hbox { mJy}$ (e.g., Diego et al. 2002). 
Both numbers indicate that we need to 
take relativistic effects into account in relating cosmological
models to survey results. Diego, Hansen \& Silk (2003) discussed how the 
corrections affect the extraction of SZE signals from {\it Planck} survey.
Here we study the impact on the SZE cluster number counts for different $w$ models.
Keeping other cosmological parameters fixed, we particularly investigate the influence 
on the $w$ and $\sigma_8$ determination. 

For the redshift distribution of SZE clusters, 
the differences of different $w$ models mainly show up at high redshifts.
To extract $w$ and $\sigma_8$, we then propose to use the total 
number of SZE clusters $N_{tot}$
and the number of clusters with redshift $z>=z_m$ (denoted as $N_{z\ge z_m}$)  
as constraints, where $z_m$ is flux dependent. 
Thus to apply our method, one needs to know the redshift range
of a cluster, but not its exact value. Therefore it provides an economic way 
to estimate cosmological parameters from large surveys, such as {\it Planck}.

To demonstrate relativistic effects clearly, we consider single-frequency measurements
with $\nu=353\hbox{ GHz}$ and the flux limit $S_{lim}=30\hbox { mJy}$. Real
{\it Planck} observations will be multi-frequency. Therefore for a particular cluster-detecting
method, realistic simulations including survey characteristics are needed to 
quantify carefully the influence of relativistic corrections. More discussions on this
regard are presented in Section 4.

Our study finds that for a survey of 
$10000 \hbox { deg}^2$, $\sigma_8$ can be constrained to a narrow range with
about $\pm 1\%$ accuracy around its central value if $1\sigma$ uncertainties are allowed
for both $N_{tot}$ and $N_{z\ge 0.5}$. The corresponding range for $w$
is $\pm 8\%$. Relativistic effects are significant in terms of locating
the central values of cosmological parameters. If they are ignored in deriving 
$\sigma_8$ and $w$ from the observed $N_{tot}$ and $N_{z\ge 0.5}$, the misplaced location of
$(\sigma_8, w)$ in $\sigma_8-w$ plane is outside the $3\sigma$ region 
around the real central ones. 

The paper is organized as follows. Section 2 contains the relevant analytical formulation.
Section 3 presents numerical results. Discussions are made in Section 4.

\section{Formulation}

\subsection{Thermal SZE with Relativistic Corrections Included}

The thermal SZE, i.e., the thermal distortion of the CMB spectrum, 
is generated through the scattering of CMB photons with thermalized hot electrons in, e.g.,  
clusters of galaxies (e.g., Sunyaev \& Zel'dovich 1980; Birkinshaw 1999;
Carlstrom, Holder \& Reese 2002). Because the temperature of electrons is 
generally much higher
than that of the CMB photons, inverse Compton scatterings are dominant,
and the distribution of photons shifts toward high-frequency side. 
In terms of the dimensionless quantity $x=h_p\nu/k_BT_{CMB}$, where $h_p$ and $k_B$
are the Planck's constant and the Boltzmann's constant, respectively, and $\nu$ 
and $T_{CMB}$ are, respectively, the frequency and the temperature of CMB photons,
in the nonrelativistic limit, the change of the CMB intensity can
be written as 
$$
{\Delta I \over I_0}=Q(x)y, \eqno (1)
$$
where $I_0=(2h_p/c^2)(k_BT_{CMB})^3/h_p^3$, 
$$
Q(x)={x^4e^x\over (e^x-1)^2}\bigg [{x\over \hbox{tanh} (x/2)} -4\bigg ], 
$$
and the Compton y-parameter
$$
y=\int n_e\sigma_T\bigg ({k_B T_{gas}\over m_e c^2}\bigg ) dl,
$$ 
where $n_e$ is the number density of hot electrons, $\sigma_T$ is the 
Thomson cross section,
$T_{gas}$ is the intracluster gas temperature,
$m_e$ is the electron mass, and $c$ is the speed of light. The integral is along
the line of sight.

In this limit, the shape of the distortion is independent of the gas temperature. For
$\nu < 218 \hbox { GHz}$, $Q(x) <0$, and the SZE shows as an absorber 
(or decrement) of CMB. 
For $\nu > 218 \hbox { GHz}$, $Q(x) >0$, and the SZE shows as a source 
(increment) of radiation.

For relativistic corrections, we use the analytical formula
of Itoh, Kohyama \& Nozawa (1998), which includes terms up to $(k_BT_{gas}/m_ec^2)^5$.
Then we have
$$
{\Delta I\over I_0}={yx^4e^x\over (e^x-1)^2}(Y_0+\theta_e Y_1+\theta_e^2 Y_2
+\theta_e^3 Y_3+\theta_e^4 Y_4), \eqno (2)
$$
where $\theta_e = k_BT_{gas}/(m_ec^2)$ and
$$
Y_0={x\over \hbox{tanh}(x/2)}-4,
$$
and $Y_1$, $Y_2$, $Y_3$ and $Y_4$ are complicated functions of $x$ 
[see Itoh et al. (1998) for detailed expressions]. 

In Figure 1 we show the single-scattering $\Delta I/I_0$ for $k_BT_{gas}=5\hbox { kev}$ and
for $k_BT_{gas}=20\hbox { kev}$. For each pair, the solid and the dashed lines
are the results without and with relativistic corrections, respectively.
It is seen that the effects
are significant for $k_BT_{gas}=20\hbox { kev}$, with about $25\%$ 
reduction of SZE signals at $\nu=353\hbox{ GHz}$. It is also noted that
the $\nu$ value for the zero thermal SZE point depends on the gas temperature,
in contrast with that of the nonrelativistic case (e.g., Birkinshaw 1999;
Rephaeli 1995). As pointed out by Itoh et al. (1998), 
their perturbative approximation is valid up to $k_BT_{gas}=15\hbox{ kev}$ 
in comparison with that from exact numerical calculations. 
At $k_BT_{gas}\sim 20\hbox{ kev}$, the approximated result
deviates from the full numerical one significantly
at $\nu > 500\hbox { GHz}$. At $\nu\le 353\hbox { GHz}$, 
however, equation (2) is still a 
good approximation even for $k_BT_{gas}\sim 20\hbox{ kev}$. 

Consider unresolved clusters, the total flux of the thermal SZE from a
cluster at redshift $z$ is
$$
S={1\over R_d^2(z)}\int \Delta I dA, \eqno (3)
$$
where $R_d(z)$ is the angular diameter distance to the cluster,
and the integral is over the surface area of the cluster.

With the isothermal assumption, equation (3) can be written as
$$
S=I_0{x^4e^x\over (e^x-1)^2}(Y_0+\theta_e Y_1+\theta_e^2 Y_2
+\theta_e^3 Y_3+\theta_e^4 Y_4)\hbox{ Y} , \eqno (4)
$$
where 
$$
Y={1\over R_d^2} \int y dA . \eqno (5)
$$

If we denote $f_{b}$ as the gas fraction that is assumed to be a constant
and $X$ as the hydrogen mass fraction, then 
$$
Y={\sigma_T \over 2m_em_pc^2}f_{b}(1+X)k_BT_{gas}MR_d^{-2},  \eqno (6)
$$
where $M$ is the total mass of the cluster, and $m_p$ is the proton mass. Further under the 
equilibrium condition, the temperature $T_{gas}$ and the mass $M$
are related through the following relation (e.g., Wang \& Steinhardt 1998)
$$
k_BT_{gas}=0.944 f_{\beta} (1+z) [\Omega_m(0)\Delta_c]^{1/3}
\times \bigg [1-{2\Omega_{\Lambda}(z)\over \Delta_c\Omega_m(z)}\bigg ]^{-3/2}
\bigg ({M\over 10^{15}h^{-1} M_{\odot}}\bigg )^{2/3}, \eqno (7)
$$
where $f_{\beta}=f_{\mu}\mu/\beta$ with $f_{\mu}$ a constant factor of
order unity that reflects the deviation from the simple
spherical model, $\mu$ the mean molecular weight of the gas,
and $\beta$ the ratio of the kinetic energy of galaxies to the thermal energy
of the gas, $\Omega_m(z)$ and $\Omega_{\Lambda}(z)$ are the matter density
parameter and the dark energy density parameter at redshift $z$, respectively,
and $\Delta_c$ is the ratio of the mean density of the cluster 
to the background matter density at redshift $z$. Here $\mu$ is related to 
$X$ by $\mu=4/(5X+3)$.

Then given a flux limit for a SZE survey, one can get the corresponding
mass limit $M_{lim}(S_{lim},z)$ from equations (4), (6) and (7). Note that the 
mass limit depends on cosmological models.

\subsection{Cluster Number Densities and Cosmological Models}

For the comoving number density of clusters, we use the fitting 
formula of Jenkins et al. (2001). Then we have
$$
{dn\over dM}(z,M)=0.315{\rho_0\over M}{1\over \sigma_M}
|{d\sigma_M \over dM}|\exp[-|0.61-\ln(D_z\sigma_M)|^{3.8}] , \eqno (8)
$$
where $\rho_0$ is the present matter density of the universe, $\sigma_M$
is the rms of the linearly-extrapolated-to-present matter density fluctuation over 
the mass scale M, and $D_z$ is the linear growth factor of density perturbations.

For $\sigma_M$, we adopt the form given by Viana \& Liddle (1999)
$$
\sigma_M=\sigma_8\bigg [ {R(M)\over 8h^{-1} \hbox{Mpc}}\bigg ]^{-\gamma[R(M)]},
\eqno (9)
$$
where $h$ is the Hubble constant in units of 
$100 \hbox { kms}^{-1} \hbox {Mpc}^{-1} $, the spatial scale $R$ 
is related to $M$ through
$$
R(M)={M\over (4\pi /3) \rho_0}, 
$$
and
$$
\gamma(R)=(0.3\Gamma+0.2)\bigg [2.92+\log_{10}
\bigg ( {R\over 8h^{-1} \hbox {Mpc}}\bigg )\bigg ], \eqno (10)
$$
with $\Gamma$ the shape parameter of the linear density fluctuation
spectrum.

In this paper, we consider flat cosmological models with the dark
energy part described by the equation of state $w=p/\rho$. 
The density parameter of dark energy is then specifically
denoted by $\Omega_Q$. The linear growth factor $D_z$ is given by 
the analytical formula of Wang \& Steinhardt (1998)
$$
D_z\approx a\exp\bigg \{ \int_a^1{da\over a}[1-\Omega_m(z)^{\alpha}]\bigg \}, \eqno (11)
$$
where $a=1/(1+z)$ is the scale factor of the universe, and
$$
\alpha \approx {3\over 5-w/(1-w)}+{3\over 125}{(1-w)(1-3w/2)\over (1-6w/5)^3}
[1-\Omega_m(z)].
$$

For the over-density of a cluster collapsed and virialized at 
redshift $z_c$, we have (Wang \& Steinhardt 1998)
$$
\Delta_c(z_c)=\zeta \bigg ({R_{ta}\over R_{vir}}\bigg )^3
\bigg ({1+z_{ta} \over 1+z_c} \bigg )^3 , \eqno (12)
$$
where $R_{ta}$ and $R_{vir}$ are the radii of the cluster at 
the time of turn-around ($z=z_{ta}$) and at the time of virialization
($z=z_c$), respectively, and
$$
\zeta={\rho_{cluster}(z_{ta})\over \rho_b(z_{ta})}
\approx \bigg ({3\pi\over 4}\bigg )^2\Omega_m^{-0.79+0.26\Omega_m-0.06w}
|_{z_{ta}}. \eqno (13)
$$
The redshifts $z_{ta}$ and $z_c$ are related through $t(z_c)=2t(z_{ta})$,
i.e., the time at $z_c$ is twice the turn-around time, and
$$
{R_{vir}\over R_{ta}}={1-\eta_v/2\over 2+\eta_t-3\eta_v/2}, \eqno (14)
$$
where
$$
\eta_t=2\zeta^{-1}{\Omega_Q(z_{ta})\over \Omega_m(z_{ta})},
$$
and
$$
\eta_v=2\zeta^{-1}\bigg( {1+z_c\over 1+z_{ta}}\bigg )^3
{\Omega_Q(z_{c})\over \Omega_m(z_{c})}.
$$




\section{Results of Analyses}

In the following studies, we take $f_{b}=0.1$, $f_{\beta}=1.4$,
and $X=0.76$ for the gas properties. For cosmological models, the
Hubble constant $H_0$ is taken to be $70 \hbox { kms}^{-1}\hbox{Mpc}^{-1}$,
$\Omega_m(0)=0.3$, and $\Omega_Q(0)=0.7$. For perturbations,
we take $\Gamma=0.2$. The observational frequency is taken to be
$\nu = 353\hbox{ GHz}$.

\subsection{ Mass limits for Flux-limited SZE Surveys}

From equations (4), (6), and (7), it is clear that the corresponding
mass limit for a flux-limited SZE survey is cosmology-dependent, 
and this dependency is one of the factors that contribute to the 
differences of the expected SZE survey results for different cosmological 
models. 

In Figure 2, we show $M_{lim}$, in units of $10^{15}h^{-1}M_{\odot}$, vs. 
redshift $z$ for different $w$ cosmologies given 
$S_{lim}=30\hbox { mJy}$. 
The three models are for $w=-1$, $w=-0.75$, and $w=-0.3$, 
respectively. 
For each pair, the lower one is the result without relativistic corrections,
and the upper one is the one with the effects included.
It is seen that the mass limit decreases as $w$ increases. 
At $z=1$, $M_{lim}(w=-0.3) \approx 0.7M_{lim}(w=-1)$.
The relative increase of the mass limit due to relativistic corrections
ranges from $6.5\%$ for $w=-1$ to $7.5\%$ for $w=-0.3$ at $z=1$.


\subsection{SZE Cluster Number Counts}

As flat models with different $w$ have different volume elements, 
linear growth factors for density perturbations, and mass limits, 
the redshift distribution of SZE clusters 
is sensitive to $w$, and thus can be used to probe the $w$ value.

In Figure 3a and Figure 3c, we show $dN/(d\Omega dz)$ for 
$S_{lim}=5\hbox { mJy}$ and $S_{lim}=30\hbox { mJy}$, respectively,
without relativistic corrections.
In each figure, results of three models with $w=-1, -0.75$ and $-0.3$
are plotted. For all the models, $\sigma_8=0.85$. 
Therefore the figures show the effects of different $w$ on the cluster redshift
distribution. 
We see that given a flux limit, models with larger $w$ (smaller 
absolute value of $w$) predict more SZE clusters over almost the
whole redshift range considered. In Figure 3b and Figure 3d, 
relativistic corrections on $dN/(d\Omega dz)$ are shown. 
It is seen that they are larger at higher redshifts, and relatively smaller
for higher $w$ at redshifts $z\ge 1$. For $S_{lim}=30\hbox{ mJy}$,
the relative changes are about $13\%$ at $z=0.5$, and at $z=2$,
they are about $41\%$, $38\%$ and $32\%$ for $w=-1, -0.75$ and $-0.3$,
respectively. 

In Figure 4, we show the surface number density $dN/d\Omega$ 
vs. $w$. For each pair, the upper and the lower ones are for the 
results without and with relativistic corrections, respectively. 
For $S_{lim}=30\hbox { mJy}$, the respective $dN/d\Omega=1.45$ and $1.30$ at $w=-0.75$,
with the change of about $10\%$.
For {\it Planck}, the survey area is about $10000\hbox{ deg}^2$.
Then the corresponding $1\sigma$ uncertainty of $dN/d\Omega$ due to Poisson fluctuations
is about $0.012$. Thus the change due to relativistic effects is about $12\sigma$. 
This indicates the significance of the effects 
in extracting $w$ value from observations.
For example, with single-frequency measurements, 
given the observed surface number density of $1.30$, 
one would infer $w\approx 0.9$ if relativistic corrections are 
neglected, which is about $20\%$ off the real value of $w=-0.75$.
We elaborate this further in the next subsection.

In Figure 3 and Figure 4, $\sigma_8=0.85$ for all the models. That is,
we assume that $\sigma_8$, as well as other cosmological parameters,
have been well determined from other observational
information. On the other hand, studies have shown that 
the total number of SZE clusters is sensitive
to $\sigma_8$, thus potentially, $\sigma_8$ and $w$ can be 
constrained from SZE surveys simultaneously. 
From the shape of the redshift distributions in Figure 3, we see that 
with the same total number of SZE clusters, the differences of the 
models are still apparent. 
For $S_{lim}=5\hbox{ mJy}$, the number of clusters with $z\ge 1$
is significantly larger for larger $w$. The ratio $dN(z\ge 1,w=-0.3)/d\Omega$
to $dN(z\ge 1,w=-1)/d\Omega$ is about $1.8$. For $S_{lim}=30\hbox{ mJy}$,
the number of clusters with $z\ge 0.5$ shows strong dependence
on $w$, with $[dN(z\ge 0.5,w=-0.3)/d\Omega] /[dN(z\ge 0.5,w=-1)/d\Omega] 
\approx 1.8$. These analyses suggest that it is possible to constrain
both $\sigma_8$ and $w$ using the total number of SZE clusters
and the number of SZE clusters with $z>z_m$ with $z_m$ flux-limit dependent.
With this methodology, we only need to know the redshift range of a cluster,
but not its precise $z$. Thus it is relatively easy to be
realized in practice. In the next subsection, we 
show how well we can apply this method to constrain $\sigma_8$ and
$w$. We also discuss relativistic effects on the parameter determination.

\subsection{Constraints on $\sigma_8$ and $w$}

We concentrate on extracting information on $\sigma_8$ and $w$ 
from SZE cluster surveys by assuming that other cosmological 
parameters have been pre-determined. The values of
those other parameters considered in this paper were listed at the beginning 
of Section 3.

Two pieces of information will be used to constrain $\sigma_8$ and $w$:
the total number of SZE clusters and the number of clusters with 
$z \ge z_m$. Here we take $S_{lim}=30\hbox { mJy}$ and $z_m=0.5$. 

The model with $\sigma_8=0.8284$ and $w=-0.75$ is adopted as the 
fiducial one. We choose $w=-0.75$ (but not $w=-1$) as our
central value so that relativistic
effects can be seen clearly (see Figure 6 in the following). Our analyzing procedures are
as follows. Given the total surface number density $dN/d\Omega$ calculated from 
the fiducial model, we search for $\sigma_8$ for
different $w$-models such that they have the same $dN/d\Omega$ as that of the 
given value. In this way, we get a relation
between $\sigma_8$ and $w$. Similarly, there is another 
$\sigma_8$-$w$ relation from the number density of clusters
with $z\ge 0.5$ [denoted as $dN(z\ge 0.5)/d\Omega$]. The two lines from the
two relations should intersect at $\sigma_8=0.8284$ and $w=-0.75$. 
Given a survey area, we can estimate the possible constrained ranges for $\sigma_8$ and $w$
allowing different uncertainty levels.

Figure 5 shows $\sigma_8$-$w$ relations for $S_{lim}=30\hbox { mJy}$ without (thin lines)
and with (thick lines) relativistic corrections.
The set of relatively flat lines are from $dN/d\Omega$, and
the other set of lines are from $dN(z\ge 0.5)/d\Omega$.
The solid lines are from the fiducial $dN/d\Omega$ and
$dN(z\ge 0.5)/d\Omega$, respectively.  The dashed lines are $\pm 3\sigma$ results
for $dN/d\Omega$, and the dotted lines are $\pm 3\sigma$ results for
$dN(z\ge 0.5)/d\Omega$. Here Poisson statistics is applied to estimate $\sigma$, and the
survey area is taken to be $10000\hbox{ deg}^2$.

For the non-relativistic case, we find that the solid lines can well be fitted by 
$$
\sigma_8(w)=0.8284|w+0.75-1|^{(0.1353-0.0150|w+0.75-1|)}, \eqno (15)
$$
and
$$
\sigma_8(w)=0.8284|w+0.75-1|^{(0.1809-0.0056|w+0.75-1|)}. \eqno (16)
$$
To avoid crowding, we did not plot $1\sigma$ and $2\sigma$ lines
in Figure 5. Our studies show that with $1\sigma$ uncertainties in
both the total number of SZE clusters and the number of clusters with 
$z\ge 0.5$, $\sigma_8$ and $w$ are constrained to be within the ranges 
$(0.8209,0.8360)$ and $(-0.81,-0.69)$, respectively. 
Thus for a survey such as {\it Planck}, 
the respective accuracies of the determined $\sigma_8$ and $w$ can possibly reach to a level of 
about $\pm 1\%$ and $\pm 8\%$. 
The $3\sigma$ determinations are about $\pm 3\%$ for $\sigma_8$ and
about $\pm 25\% $ for $w$ if other 
cosmological parameters have already been known to good precisions. 


In the relativistic case, the approximated
analytical relations between $\sigma_8$ and $w$ for the solid lines
are
$$
\sigma_8(w)=\sigma_8(-0.75)|w+0.75-1|^{(0.1246-0.0115|w+0.75-1|)}.\eqno (17)
$$
and
$$
\sigma_8(w)=\sigma_8(-0.75)|w+0.75-1|^{(0.16969-0.00009|w+0.75-1|)}. \eqno (18)
$$
The relations are slightly flatter than those of the
nonrelativistic case (eq. [15] and eq. [16]). As seen in Figure 3d, 
relativistic effects are stronger at higher redshifts. 
It is also known that models with larger $w$ (lower absolute $w$ value)
predict more high redshift clusters,  
and thus the relativistic corrections to $dN/d\Omega$ and $dN(z\ge 0.5)/d\Omega$ are larger.
This results the decrease of the differences between different
$w$ models in terms of the two measurements, and therefore the flatter 
$\sigma_8$-$w$ relations. The $1\sigma$ and $3\sigma$ 
ranges for $\sigma_8$ are $(0.8213,0.8356)$ and $(0.8056,0.8491)$, 
respectively, and the respective ranges for $w$ are $(-0.815,-0.69)$ and $(-0.95,-0.57)$. 
We see that relativistic corrections do not affect the precisions of the parameter
determination very much. 

On the other hand, however, if relativistic effects were neglected 
in extracting $\sigma_8$ and $w$ from a SZE survey, the determined 
central values can be significantly off the real ones.
To estimate this misplacement, we take the relativistic $dN/d\Omega$ and
$dN(z\ge 0.5)/d\Omega$ from the fiducial model with $\sigma_8=0.8284$ and $w=-0.75$ as inputs, 
but search for $\sigma_8$-$w$ relations with {\it nonrelativistic} mass limits. 
In Figure 6, we plot the results as solid lines. The contours are the 
$1\sigma$ (solid), $2\sigma$ (dotted) and $3\sigma$ (dashed) lines around 
$\sigma_8=0.8284$ and $w=-0.75$ (denoted by the star). Note that the solid contour is composed by 
$\pm 1\sigma$ lines for the total number of SZE clusters and the 
$\pm 1\sigma$ lines for the number of clusters with $z\ge 0.5$. The $2\sigma$ and
$3\sigma$ contours are similar. First let us consider one parameter case 
(given the other) using the
total number of SZE clusters as the constraint. Given $\sigma_8=0.8284$,
the derived central $w$ from the wrong solid line is $w\approx 0.92$,
while the $3\sigma$ value is $w\approx 0.78$. On the other hand, if 
$w=-0.75$ is given, the derived $\sigma_8$ is $\sigma_8\approx 0.814$,
while the $3\sigma$ value is $\sigma_8\approx 0.824$. We see that
the misplacement is dramatically large. 
If both $\sigma_8$ and $w$ are considered as unknown, the derived
values from the cross-point of the two solid lines are
$\sigma_8\approx 0.819$ and $w\approx -0.81$. Thus the derivations from the
fiducial values are about $1.1\%$ for $\sigma_8$ and $8\%$ for $w$.
Although this offset is smaller than
those of the one parameter cases, the derived "central values"
are outside the $3\sigma$ region (the region surrounded by the dashed contour).
Therefore relativistic effects are very significant and cannot be ignored. 






\section{Discussion}


As the most important property of SZE is its redshift-independence,
SZE cluster surveys are one of the best to probe the structure
formation at high redshifts. Because of the cosmological 
model dependence of the geometry of the universe, of the evolution 
of density fluctuations, and of the mass limit corresponding
to a given flux limit, the redshift distribution of SZE clusters
is sensitive to cosmological parameters. 

We particularly discussed the effects of relativistic corrections of
thermal SZE on the redshift distribution of clusters, 
and further on $\sigma_8$ and $w$ determinations from SZE cluster surveys. 
The corrections affect the cluster redshift distribution through changing the mass
limit for a given flux limit. The changes depend on the observing
frequency and on the flux limit. For $\nu=353\hbox{ GHz}$ and $S_{lim}=30\hbox{ mJy}$, 
at $z=1$, the relative increase of the mass limit due to relativistic effects 
is about $6.5\%$ and $7.5\%$ for $w=-1$ and $w=-0.3$, respectively. 
With $\sigma_8=0.85$ for all the models, the corresponding decrease of 
$dN/d\Omega$ is about $8.9\%$ and $13.5\%$ for $w=-1$ and $w=-0.3$, respectively. 
For $dN(z\ge 0.5)/d\Omega$, the percentages of the decrease are, respectively, 
about $18.4\%$ and $21.3\%$ for $w=-1$ and $w=-0.3$.

With fixed other cosmological parameters, we studied the constraints on $\sigma_8$ and $w$
from the total number of SZE clusters and from the number of
clusters with $z\ge z_m$, where $z_m$ is flux-limit dependent. 
For $S_{lim}=5\hbox { mJy}$, $z_m=1.0$ is appropriate. For
$S_{lim}=30\hbox { mJy}$, $z_m=0.5$. The nice part of our methodology discussed here
is that we do not need to know the precise redshift of a cluster, but 
only its range, i.e., larger or smaller than $z_m$. Therefore
it is applicable in large surveys, such as {\it Planck}, because then one can use not-so-precise
methods to estimate the redshifts of clusters, such as the photometric method
or morphological redshift estimates (Diego et al. 2003). 
Our study showed that at $S_{lim}=30\hbox { mJy}$, 
for a survey of $10000\hbox{ deg}^2$, $\sigma_8$ can be determined to a level of $\pm 1\%$
and $w$ to an accuracy of $\pm 8\%$ with $1\sigma$ uncertainties 
in both the total number of clusters and the number of high
redshift clusters ($z\ge 0.5$ in our analyses). With $3\sigma$ uncertainties,
the determined parameter ranges are $\pm 3\%$ and $\pm 25\%$ around the
central values of $\sigma_8$ and $w$, respectively. In our uncertainty estimations
in subsection 3.3, we assumed that the redshift information is available for $100\%$ SZE clusters.
That is, when we assessed the $\sigma$ for the number of clusters with $z\ge 0.5$, 
we used the full number of clusters with $z\ge 0.5$ for a survey of $10000\hbox{ deg}^2$.
If the redshifts are available for $10\%$ of clusters, with $1\sigma$ uncertainties,
the determined $\sigma_8$
and $w$ have the ranges $(0.815,0.842)$ and $(-0.89,-0.62)$ around the central
values $\sigma_8=0.8284$ and $w=-0.75$. Thus with $10\%$ samples having redshift
information, the accuracies of $1\sigma$ determination for $\sigma_8$ and $w$ are 
$\pm 1.6\%$ and $\pm 18\%$, respectively.
Relativistic corrections do not have significant effects in this aspect.

However, we found that the central values of $\sigma_8$ and $w$ can be
significantly misplaced if relativistic corrections are not included
in extracting cosmological information from observed cluster number counts.
For $S_{lim}= 30\hbox{ mJy}$, the offset for $\sigma_8$ is about $1.1\%$, and about $8\%$
for $w$. In terms of the percentages, this misplacement is comparable to
the $1\sigma$ deviation. But the location of the wrong central values in the
$\sigma_8-w$ plane is significantly outside the $3\sigma$ region around the
real central $(\sigma_8, w)$ as seen in Figure 6. Therefore, relativistic effects
are important, and should be considered carefully in analyzing survey results.


We need to emphasize that our discussions on relativistic effects concentrate on
single-frequency measurements at $353 \hbox{ GHz}$. In reality, {\it Planck} will conduct
multi-frequency observations, and the overall relativistic effects could be weaker than the
case of $\nu=353 \hbox{ GHz}$ depending on the techniques used to detect SZE clusters.
Diego, Hansen \& Silk (2003) studied the impact of relativistic corrections on the reconstruction
of the $y$ map from simulated $\Delta T/T$ of different frequency channels
with Bayesian non-parametric method. They found that for clusters of $T=10\hbox{ kev}$,
the relative difference in the recovered $y$ between
relativistic and non-relativistic approaches in the component separation was about
$4\%$ for low-redshift clusters and about $8\%$ for high-redshift clusters. This difference is
smaller than that at $\nu=353 \hbox{ GHz}$, which is about $15\%$. This is because the
$y$ maps were constructed through the weighted average of $\Delta T/T$ of 
different frequency channels, 
and relativistic corrections are smaller than $10\%$ for $\nu < 353 \hbox{ GHz}$.
Thus the single-frequency estimation of relativistic effects presented in this paper
can be an over-estimation. To quantify the impact realistically, one can, with simulated
maps including characteristics of a survey, get two sets of mass limits of
detectable SZE clusters at different redshifts with
and without relativistic corrections for a particular cosmological model.
Then analytical studies similar to those shown in this paper can be carried out based
on the two sets of mass limits.
We would like to mention that the flux limit for completeness, thus the mass limit,
depends on the method applied to find clusters. The limit $S_{lim}=30\hbox{ mJy}$
(at $\nu=353 \hbox{ GHz}$) adopted in our analyses is from the estimation of Diego et al. (2002)
based on the Maximum Entropy Method (Hobson et al. 1998; Hobson et al. 1999). On the
other hand, Bayesian non-parametric method gives the completeness limit
$S_{lim}\sim 200\hbox{ mJy}$ (at $\nu=353 \hbox{ GHz}$) (Diego et al. 2003),
and a multifilter approach of Herranz et al. (2002) finds $S_{lim}\sim 170\hbox{ mJy}$.
With such high flux limits, relativistic corrections can be large even
with the multi-frequency diminishing effect taken into account.

The constraints on $\sigma_8$ and $w$ shown in the paper are
derived under the assumption that all other cosmological parameters have been
determined by other studies. If we allow, e.g., $\Omega_m(0)$, to vary as well, 
the estimation can be affected considerably (e.g., Haiman, Holder, \& Mohr 2001).
On the other hand, however, there are indeed different types of observations that can give good
constraints on those parameters. For example, galaxy surveys, such as two degree
field Galaxy Redshift Survey (2dFGRS) (Colless et al. 2001) and the Sloan Digital
Sky Survey (SDSS), can put constraints on the shape of the 
perturbation spectrum, which primarily depends on $\Omega_mh$ (Szalay et al.
2001; Dodelson et al. 2002; Percival et al. 2001; 2002). Combined with
the measurements on the Hubble constant $H_0$, $\Omega_m$ can be 
well constrained. In any case, studies on the multiple-parameter estimation 
with the methodology put forward in this paper and on the
impact of relativistic corrections are desired, and will be
carried in the future.



\acknowledgments

We thank the referee for the helpful comments and suggestions.
This research was supported in part by the National Science Foundation of China
under grant 10243006, and by the Ministry of Science and Technology of China under
grant TG1999075401.

\clearpage



\begin{figure}
\plotone{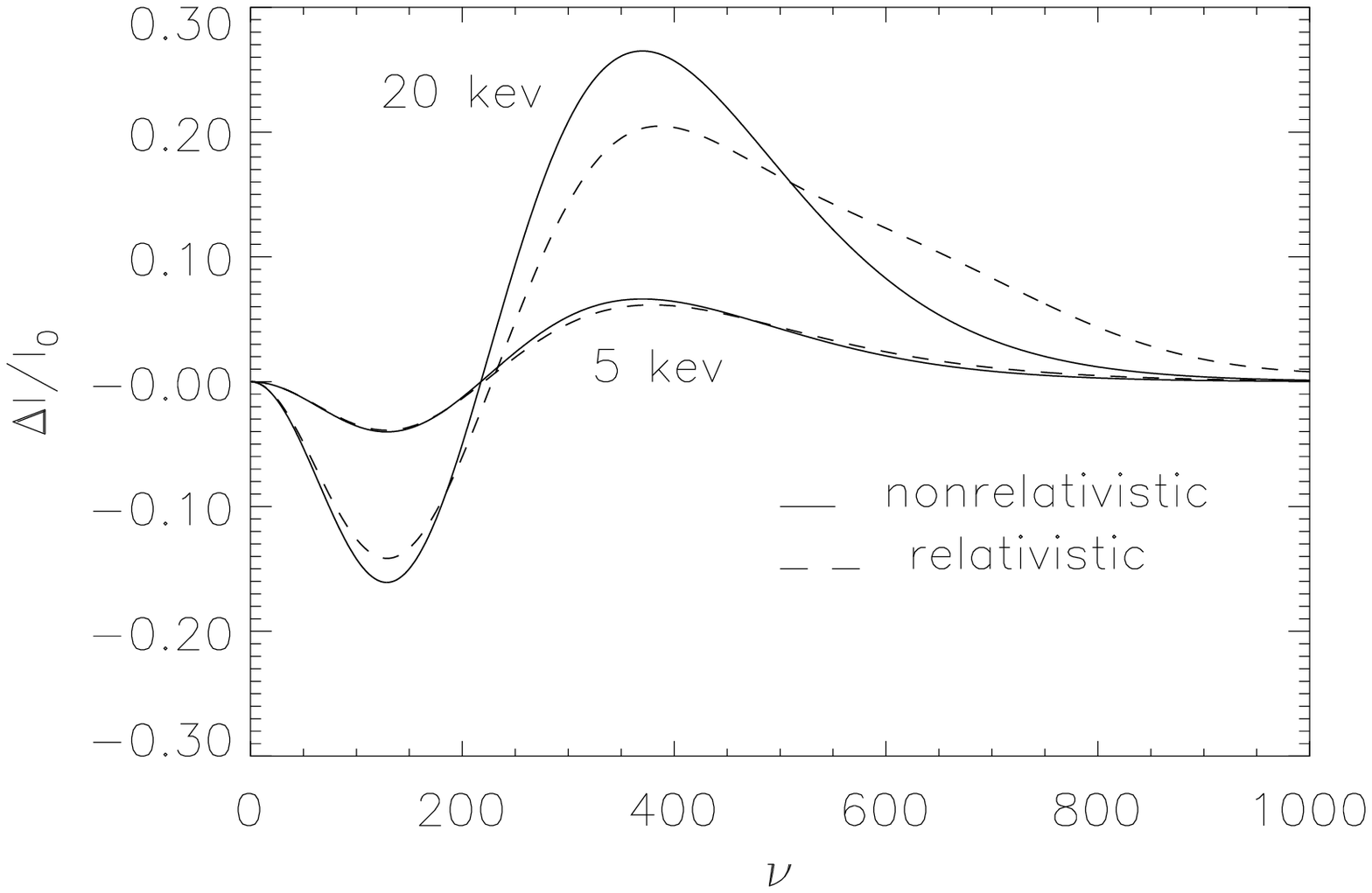}
\caption{Thermal SZE $\Delta I/I_0$ after a single scattering 
with $I_0=(2h_p/c^2)(k_BT_{CMB})^3/h_p^3$
vs. frequency $\nu$ for $k_BT_{gas}=5\hbox{ kev}$ and $20\hbox{ kev}$.
In each pair, the solid line represents the result without
relativistic corrections, and the dashed line is for the results
with relativistic corrections up to $(k_BT_{gas}/m_ec^2)^5$.
\label{f1}}
\end{figure}

\begin{figure}
\plotone{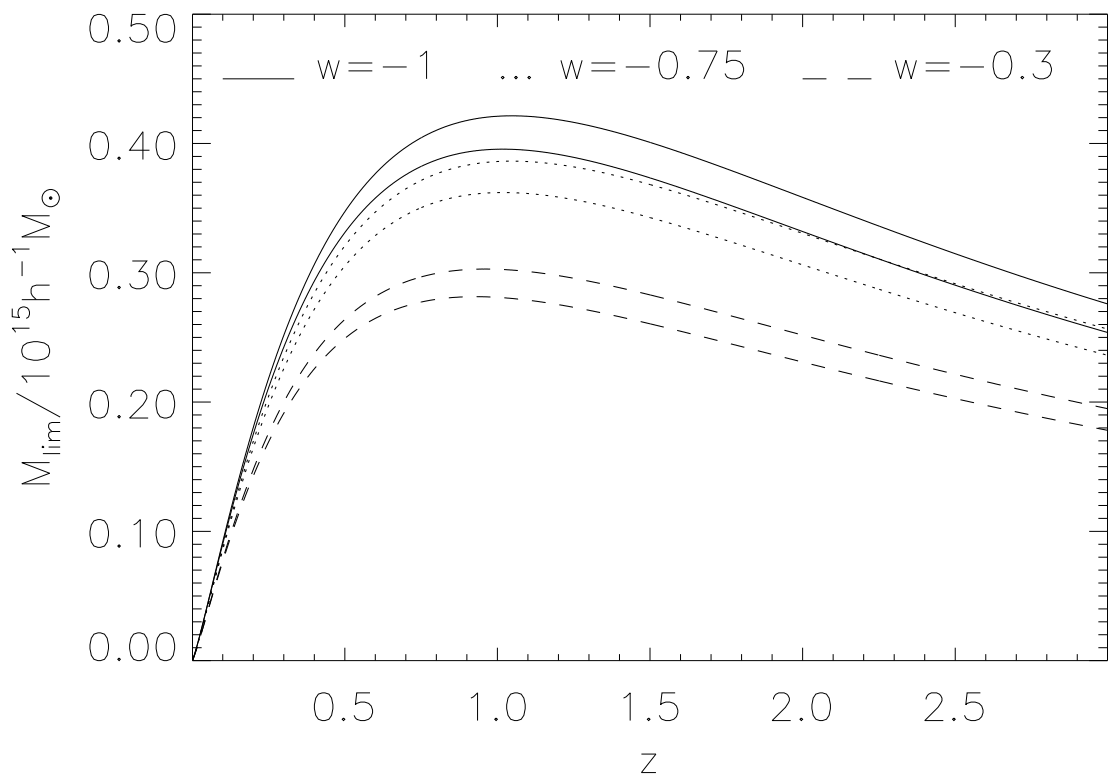}
\caption{The mass limit $M_{lim}$ in units of $10^{15}h^{-1}M_{\odot}$ vs. redshift $z$.
The flux limit is $S_{lim}=30\hbox{ mJy}$. The solid lines are for $w=-1$,
the dotted lines are for $w=-0.75$ and the dashed lines are for $w=-0.3$.
In each pair, the upper one corresponds to the one with relativistic corrections
included, and the lower one is for nonrelativistic $M_{lim}$.
\label{f2}} 
\end{figure}



\begin{figure}
\plotone{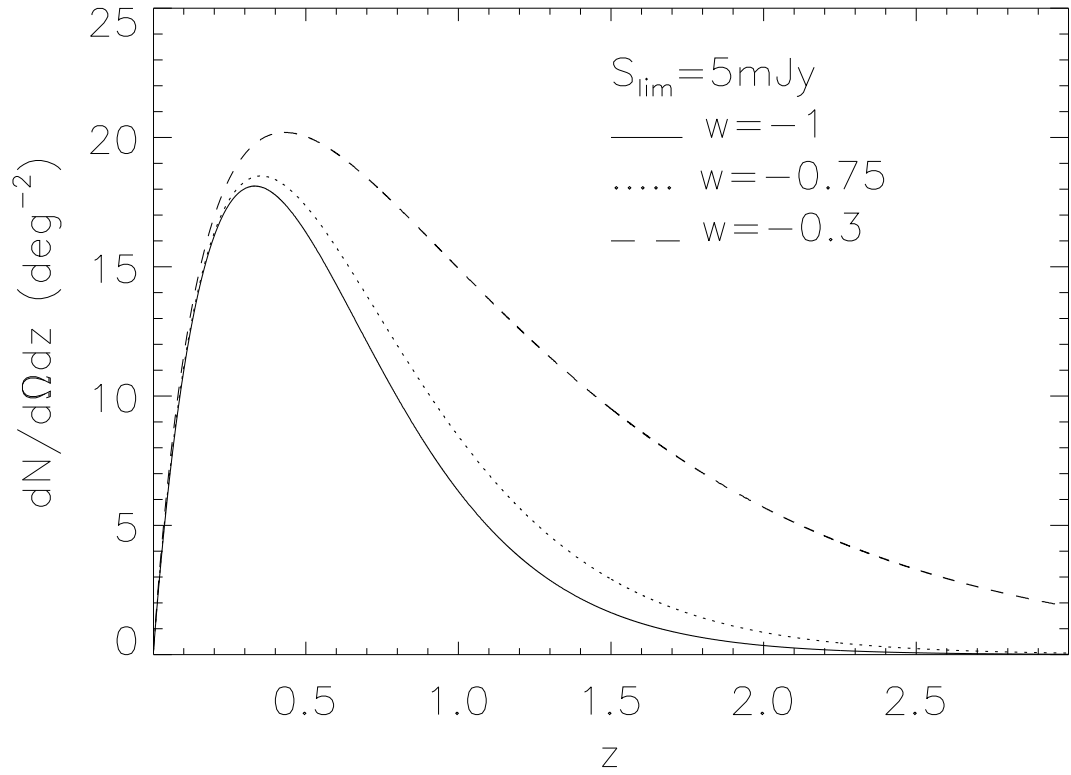}
\caption{The redshift distribution of SZE clusters $dN/(d\Omega dz)$ 
in units of $\hbox{ deg}^{-2}$ for
$S_{lim}=5\hbox { mJy}$. No relativistic
corrections are included, and $\sigma_8=0.85$.
The solid line is for $w=-1$, the dotted line is for $w=-0.75$, and
the dashed line is for $w=-0.3$.
\label{f3a}} 
\end{figure}

\begin{figure}
\plotone{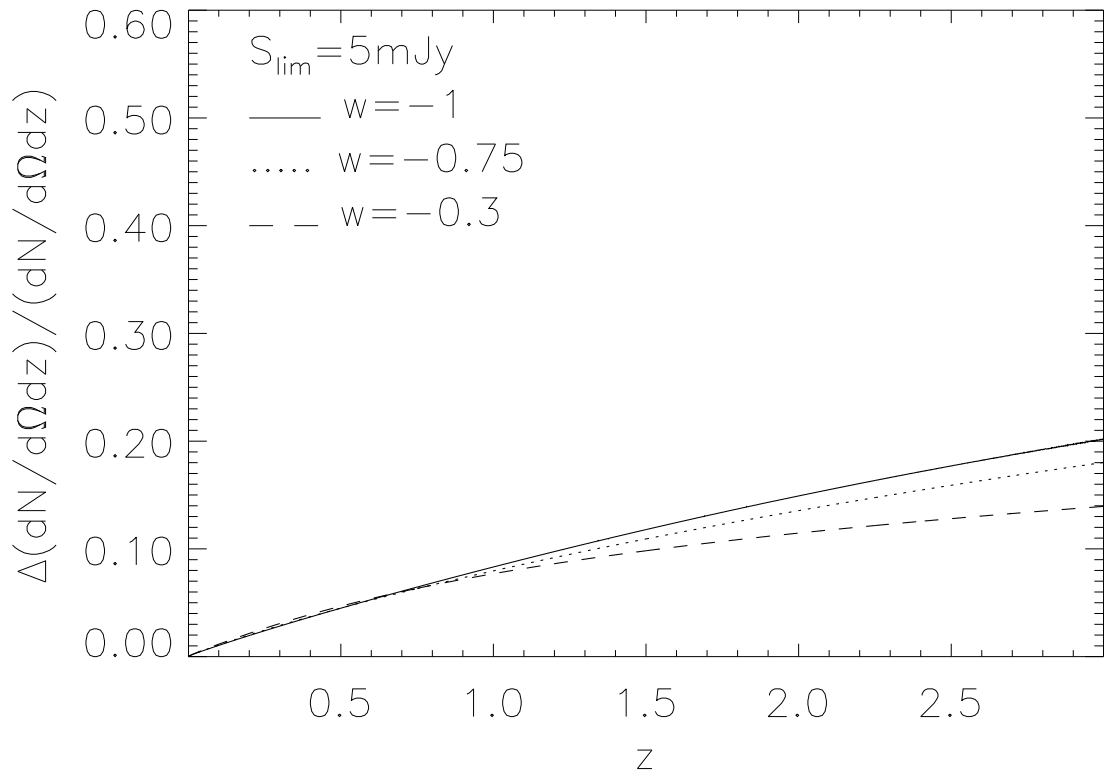}
\caption{Relative changes of the redshift distribution of SZE clusters
due to relativistic corrections for $S_{lim}=5\hbox { mJy}$ and $\sigma_8=0.85$.
\label{f3b}}
\end{figure}

\begin{figure}
\plotone{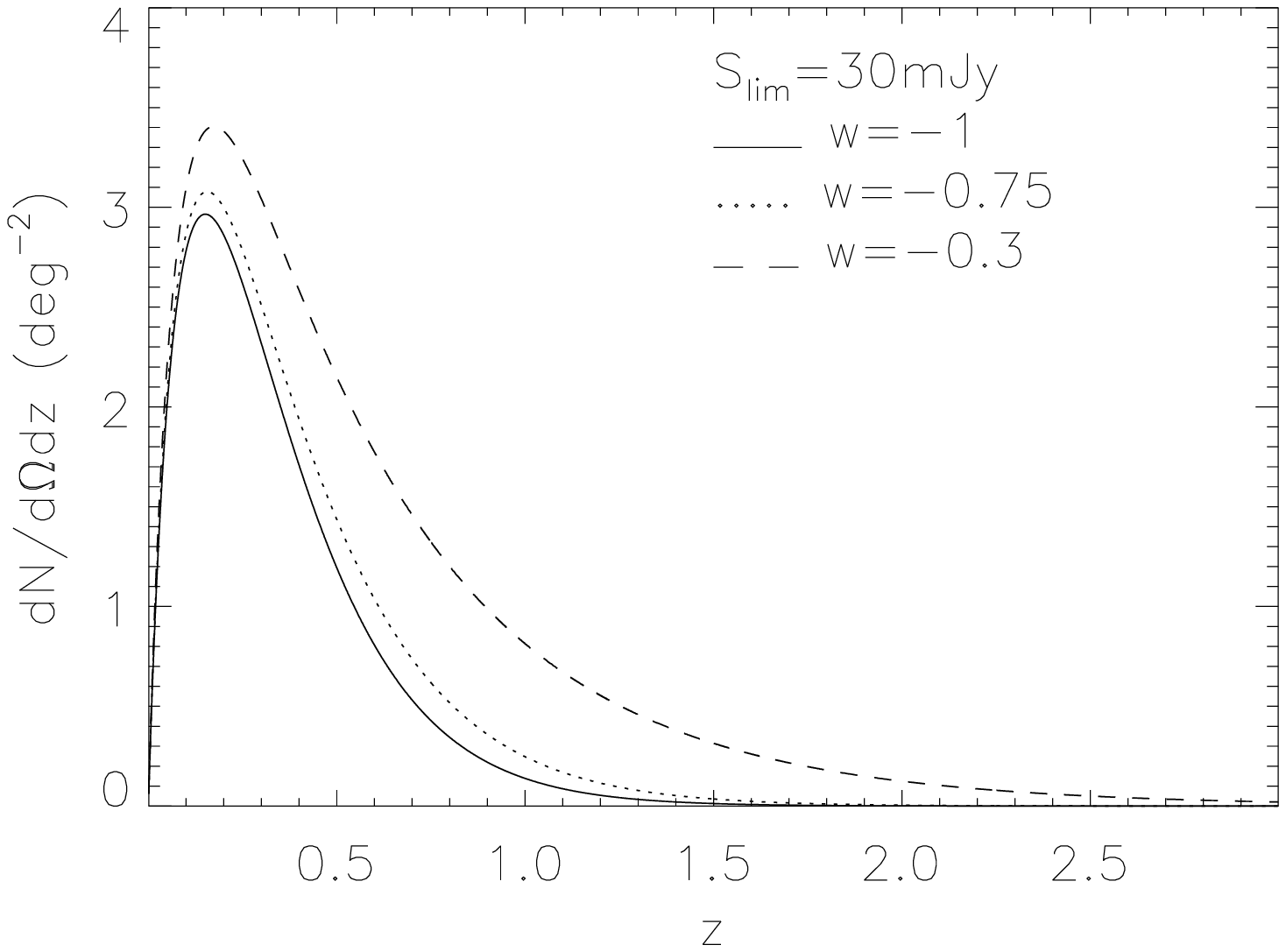}
\caption{Same as fig.3a but for $S_{lim}=30\hbox { mJy}$
\label{f3c}}
\end{figure}

\begin{figure}
\plotone{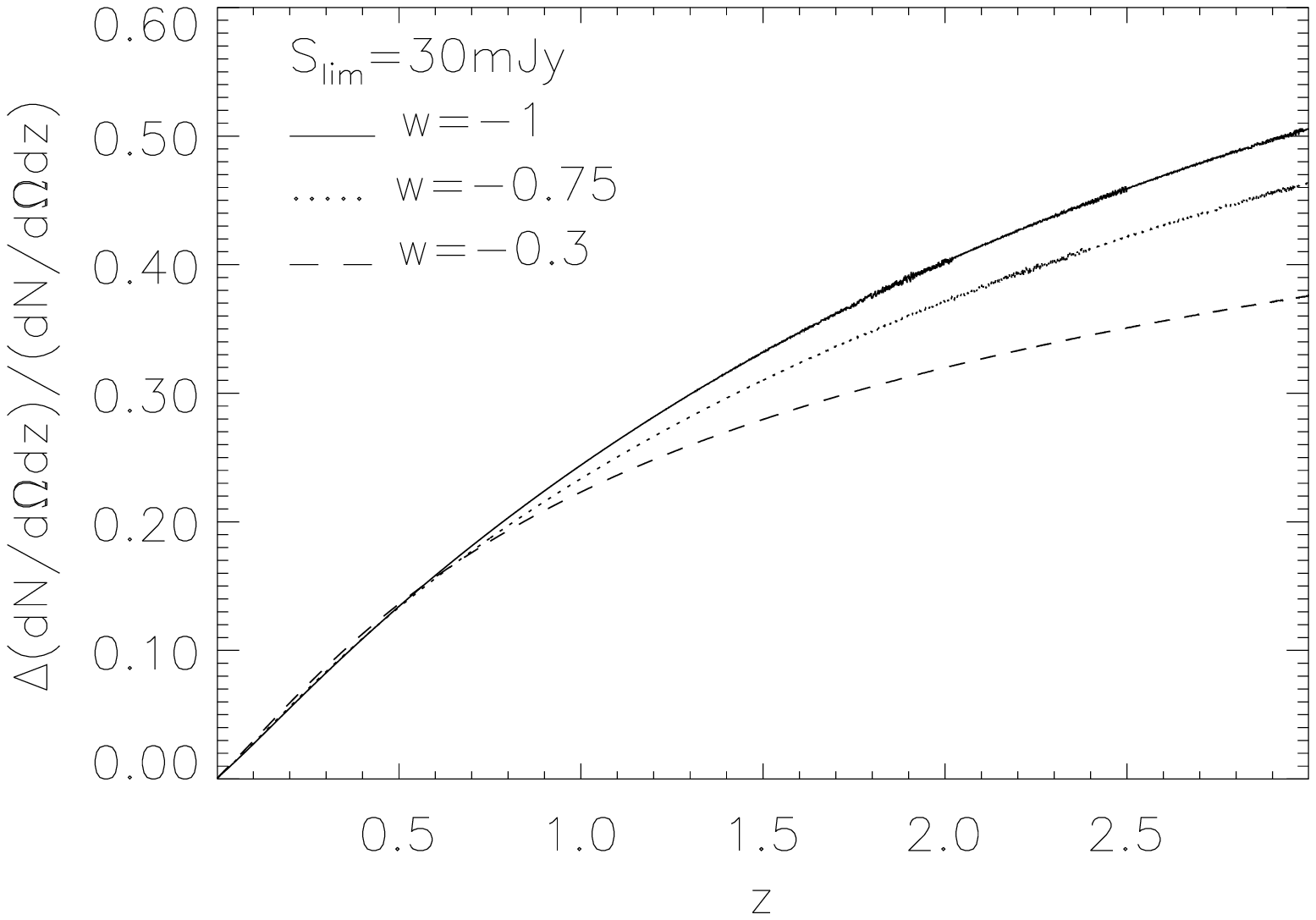}
\caption{Same as fig.3b but for $S_{lim}=30\hbox { mJy}$
\label{f3d}}
\end{figure}

\begin{figure}
\plotone{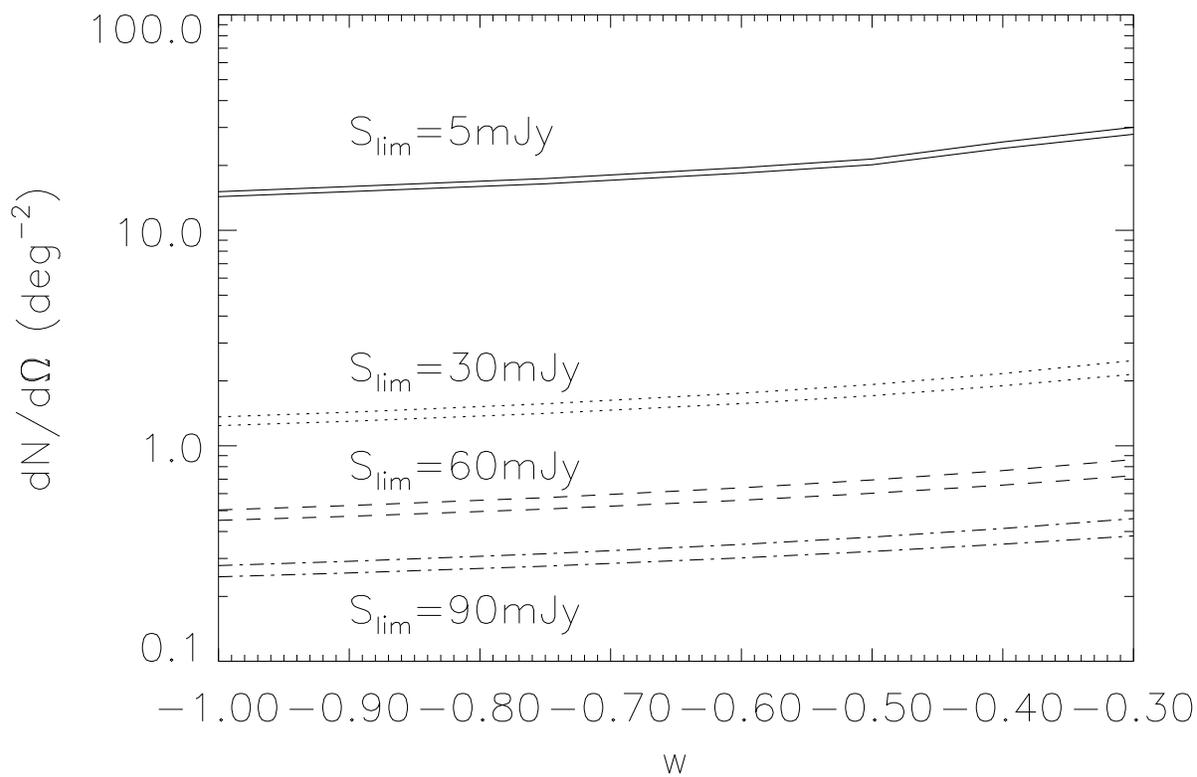}
\caption{The surface number density of SZE cluster $dN/d\Omega$ in units of $\hbox{ deg}^{-2}$ vs.
$w$. The solid, dotted, dashed and dash-dotted lines are for $S_{lim}=5\hbox { mJy}$,
$30\hbox { mJy}$, $60\hbox { mJy}$ and $90\hbox { mJy}$, respectively. In each pair,
the upper one is the nonrelativistic result, and the lower one is
the result with relativistic corrections included. Here $\sigma_8=0.85$ for all the
models.
\label{f4}} 
\end{figure}


\begin{figure}
\plotone{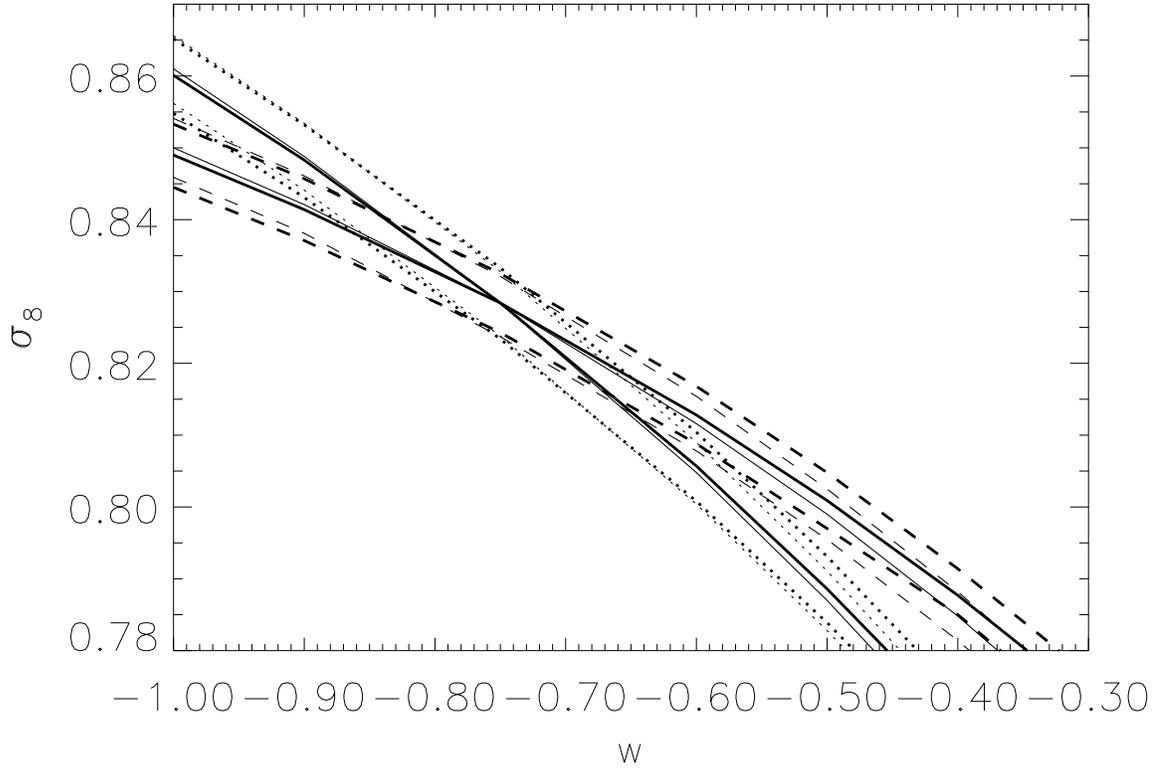}
\caption{The $\sigma_8-w$ relations without (thin lines) and with (thick lines)
relativistic corrections. The set of relatively flatter lines
are from the total number of SZE clusters, and the steeper lines
are from the number of clusters with $z\ge 0.5$.
The solid lines are from the central values of $N_{tot}$ and $N_{z\ge 0.5}$,
respectively. The dashed lines are the $\sigma_8-w$ relations from
$N_{tot}\pm 3\sigma$, respectively,
and the dotted lines are the respective relations from $N_{z\ge 0.5}\pm 3\sigma$.
The survey area is taken to be $10000\hbox{ deg}^2$.
\label{f5}} 
\end{figure}


\begin{figure}
\plotone{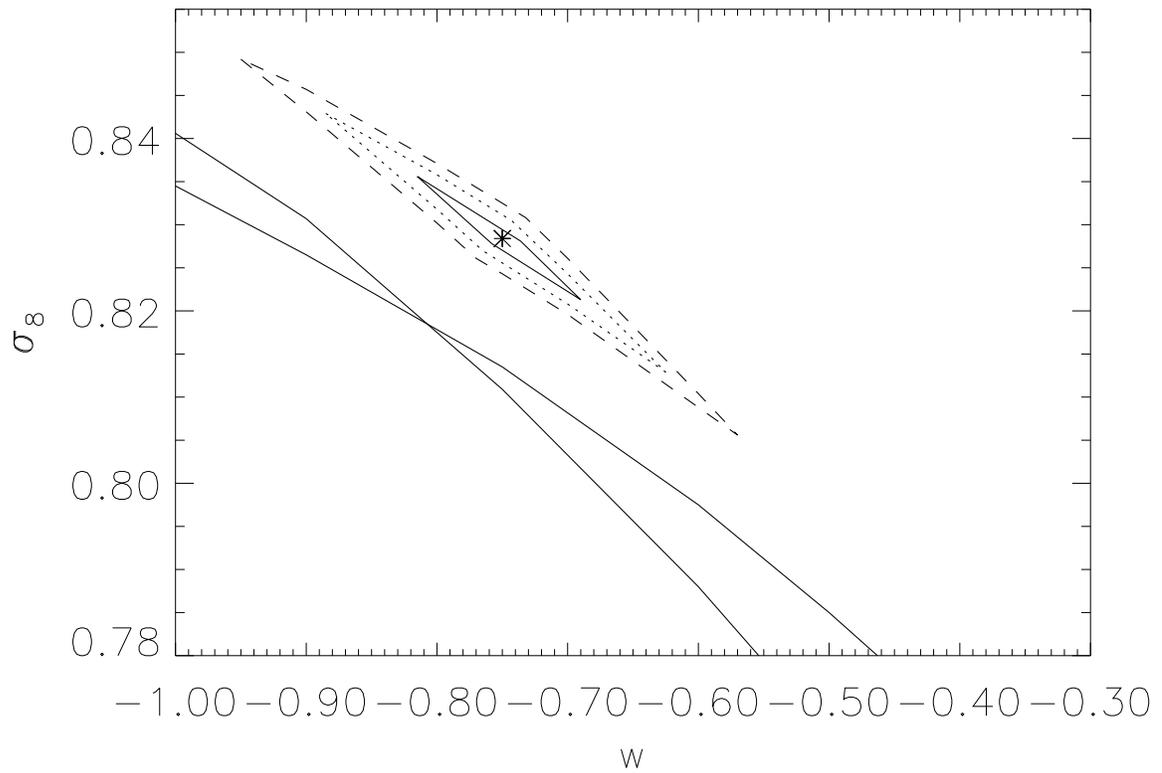}
\caption{The misplacement of the central $\sigma_8$ and $w$. The two solid lines
are from the nonrelativistic calculations given $N_{tot}$ and
$N_{z\ge 0.5}$ calculated relativistically with $w=-0.75$
and $\sigma_8=0.8284$ (represented by the star in the figure). The
survey area is taken to be $10000\hbox{ deg}^2$.
The solid, dotted, and dashed closed lines are $1\sigma$, $2\sigma$
and $3\sigma$ deviations around $w=-0.75$ and $\sigma_8=0.8284$ (see the text for details).
\label{f6}} 
\end{figure}

\clearpage






\end{document}